# A WSe$_2$ vertical field emission transistor

*Antonio Di Bartolomeo*[1,2,3,*], *Francesca Urban*[1,2,3], *Maurizio Passacantando*[4], *Niall McEvoy*[5], *Lisanne Peters*[5], *Laura Iemmo*[1,2,3], *Giuseppe Luongo*[1,2,3], *Francesco Romeo*[1,3], *and Filippo Giubileo*[3]

[1] Physics Department "E. R. Caianiello", University of Salerno, via Giovanni Paolo II n. 132, Fisciano 84084, Italy

[2] Interdepartmental Centre NanoMates, University of Salerno, via Giovanni Paolo II n. 132, Fisciano 84084, Italy

[3] CNR-SPIN Salerno, via Giovanni Paolo II n. 132, Fisciano 84084, Italy

[4] Department of Physical and Chemical Science, University of L'Aquila, and CNR-SPIN L'Aquila, via Vetoio, Coppito 67100, L'Aquila, Italy

[5] AMBER & School of Chemistry, Trinity College Dublin, Dublin 2, Ireland

*E-mail: adibartolomeo@unisa.it

*Abstract*

We report the first observation of gate-controlled field emission current from a tungsten diselenide (WSe$_2$) monolayer, synthesized by chemical-vapour deposition on SiO$_2$/Si substrate. Ni contacted WSe$_2$ monolayer back-gated transistors, under high vacuum, exhibit n-type conduction and drain-bias dependent transfer characteristics, which are attributed to oxygen/water desorption and drain induced Schottky barrier lowering, respectively. The gate-tuned n-type conduction enables field emission, i.e. the extraction of electrons by quantum tunnelling, even from the flat part of the WSe$_2$ monolayers. Electron emission occurs under an electric field ~100 V μm$^{-1}$ and exhibit good time stability. Remarkably, the field emission current can be modulated by the back-gate voltage. The first field-emission vertical transistor based on WSe$_2$ monolayer is thus demonstrated and can pave the way to further optimize new WSe$_2$ based devices for use in vacuum electronics.

*Introduction*

In the past decade, transition metal dichalcogenides (TMDs) have attracted a lot of attention due to several promising properties for electronic and optoelectronic applications, including pristine interfaces without out-of-plane dangling bonds, flexibility, tuneable bandgap in the range 0.7-2.2 eV, moderate mobility comparable to that of Si channels in modern devices, transparency, broadband light response, photoluminescence, thermal stability in air and high scalability in device fabrication [1–3]. Mechanical and liquid exfoliation [4], chemical vapour deposition (CVD) [5,6], pulsed laser deposition [7] molecular beam epitaxy [8,9], etc. have been used for their production. Most of the TMDs, such as MoS$_2$, MoSe$_2$, WS$_2$, MoTe$_2$, HfSe$_2$, ZrSe$_2$, show natural n-type conduction, with p-type behaviour achieved only through suitable doping or coupling to other conducting or dielectric materials [10,11]. On the contrary, air exposed WSe$_2$, that is an important member of the TMD family, shows predominantly p-type conduction or ambipolar behaviour [12–15]. Similar to other TMDs, the WSe$_2$ band structure evolves from a narrow ($\sim 1.2\ eV$) and indirect bandgap in the bulk form to a wider ($\sim 1.6\ eV$) and direct bandgap in the monolayer counterpart [16–18]. Besides the number of layers, a vertical electric field can enable further bandgap tuning [19]. WSe$_2$ has also emerged among other TMDs, and in particular over the widely studied MoS$_2$, due to its lower electron and hole effective masses and thus higher mobility, demonstrated up to the record value of $500\ \frac{cm^2}{Vs}$ for holes at room temperature [13,20,21]. The high bandgap and mobility make WSe$_2$ a suitable two-dimensional (2D) channel material for high speed flexible electronics [22–24]. WSe$_2$ complementary inverters [25] and field effect transistors (FETs) with the ideal subthreshold swing $SS = 60\ \frac{mV}{decade}$ [26] have been demonstrated. Furthermore, the strong light interaction, with white light absorption of 5-10% in a monolayer [27], and the giant luminescence at room temperature [28] have been exploited in WSe$_2$-based photodetectors [29,30] or light emitting diodes [31].

As for any other semiconducting materials, the control of the doping type and level in WSe$_2$ is an important parameter for its electronic and optoelectronic exploitation and, to date, chemical and substitutional doping are the two major applied doping strategies. The first approach consists of the adsorption of atoms or molecules on the surface of the material, which leads to the alteration of its electronic structure as a consequence of surface charge transfer [26,32], while substitutional doping of the transition metal or of the chalcogen atoms is pursued in the second approach [33,34]. Electric control of carrier concentration by a gate in field-effect transistor structures is another effective way to tune the doping level and type. Such strategy is viable, for instance, in field emission (FE) applications, where electrons are extracted from the material by quantum tunnelling through the surface potential barrier upon the application of a strong electric field. FE is of great relevance to a variety of applications, ranging from electron microscopy and lithography to display technology or vacuum electronics. Indeed, lateral field emission transistors have recently gained popularity due to their potential for many high-frequency and high-power applications [35,36]. In such devices, the traditional approach of placing a gate roughly in between the source (cathode) tip and the drain (anode), has been replaced by a gate behind the emitting tip, controlling its doping level and conductivity. In such a way, a narrower gate voltage can modulate the source-drain field-emission current by depleting or enhancing the carrier density available at the source for tunnelling [37]. The gate control of doping and therefore of the FE current is particularly effective when the cathode is made of a low-dimensional material. Indeed, FE current tuning up to six orders of magnitude, by a modest gate voltage up to 20V, has been reported for graphene vertical emitters [38]. Also, high emission current density over $0.1\ Acm^{-2}$, low turn-on field of few $\frac{V}{\mu m}$ and sensitive gate modulation were obtained from micro-gated graphene nanomesh field emitter arrays fabricated on a SiO$_2$/Si substrate, used as the back gate [39]. The increasing positive gate was observed to lower the turn-on field and increase the field-enhancement factor. Moreover, electron FE in field-effect transistors with vacuum transport parallel to the back-gate substrate have been reported from planar graphene edge sources in which the field emitted electrons have been shown to have an energy spectrum very near the Fermi energy of the graphene source [40,41].

Despite the suitable combination of mechanical and electronic properties, to date, only few studies have considered TMDs in field emission devices. TMDs possess atomically sharp edges like graphene, as well as localized defects such as vacancies or substitutional atoms, which can enhance the surface electric field and lower the on-set voltage needed to enhance the tunnelling probability of electrons to vacuum. FE current from $MoS_2$ flakes with low turn-on field and high field enhancement factor has been reported both from the edges [42] and the flat part [43,44] of few-layers $MoS_2$ flakes. The natural n-doping of $MoS_2$, possibly controlled by a back gate, has been pointed out as an important ingredient for future applications of $MoS_2$ FE in vacuum nanoelectronic and flat panel displays.

The recent observation of metallic edges in atomically thin $WSe_2$ monolayers grown by CVD [45], the lower bandgap (∼1.6 vs. ∼1.8 eV), effective electron mass (0.33 vs. 0.57 $m_0$, the rest mass of the electron) and electron affinity (∼3.9 vs. ∼4.2 eV) [46] would suggest that $WSe_2$ is a far better field emitter than $MoS_2$. Yet, no field emission has been reported from $WSe_2$ layers to date.

In this paper, we use CVD to fabricate monolayer and few-layer $WSe_2$ flakes on $SiO_2$/Si substrate and investigate their electrical properties, in high vacuum, using back-gated transistor structures. We show that the $WSe_2$ flakes, contacted by Ni, exhibit n-type conduction, with conductivity highly controlled by the back-gate voltage. Taking advantage of the gate-controlled n-type doping, we locally probe the FE current from a monolayer $WSe_2$ and we achieve a FE current in the range on the $\mu A$ from the flat part of the flake. More importantly, we demonstrate that the FE current can be modulated by the back-gate voltage, thus realizing the first vertical FE transistor based on a $WSe_2$ monolayer. We unveil the physics mechanisms underlying the operation of such a device and give indications for its optimization to enhance its driving current capability and to lower the applied voltage. This study can pave the way to the further exploitation of $WSe_2$ in a new generation of devices for vacuum electronics.

*Experimental*

$WSe_2$ flakes, consisting of mono and few layers, were synthesized by CVD on heavily doped p-Si substrates (Boron doped, resistivity $< 0.005\ \Omega\ cm$), covered by 300 nm dry thermal $SiO_2$. The CVD growth was performed in a two-zone quartz tube furnace. The growth substrate was placed face down on a WOx seed layer, which was made by sputtering a 20 nm layer of W, using a GATAN 682 Precision Etching and Coating System (PECS), followed by an oxidation step on a hotplate for 1 h at 500 °C. The seed layer-substrate stack creates a microreactor, similar to previous reports on $MoS_2$ growth [47]. This microreactor was placed in the centre of the high-temperature zone of the furnace, heated to 850 °C. Se was placed upstream in the low-temperature heating zone, which was independently heated to 250 °C. A 50 sccm flow of Ar:$H_2$ (9:1) transported Se vapour from the low-temperature zone to the high-temperature zone where it reacted with the WOx in the microreactor. A reaction of 40 minutes and a growth pressure of 6.0±0.2 Torr were used to grow the $WSe_2$ flakes, which, initially selected by optical microscopy, were characterized by Raman and photoluminescence (PL) spectroscopy. Some flakes, identified as monolayers, were used for the fabrication of FETs with Ni/Au ($5/50\ nm$) contacts. The Ni/Au contacts were patterned by electron beam lithography in a sequence of parallel leads for two or four probe measurements. The metals were deposited by sputtering using a Gatan precision etching and coating system.

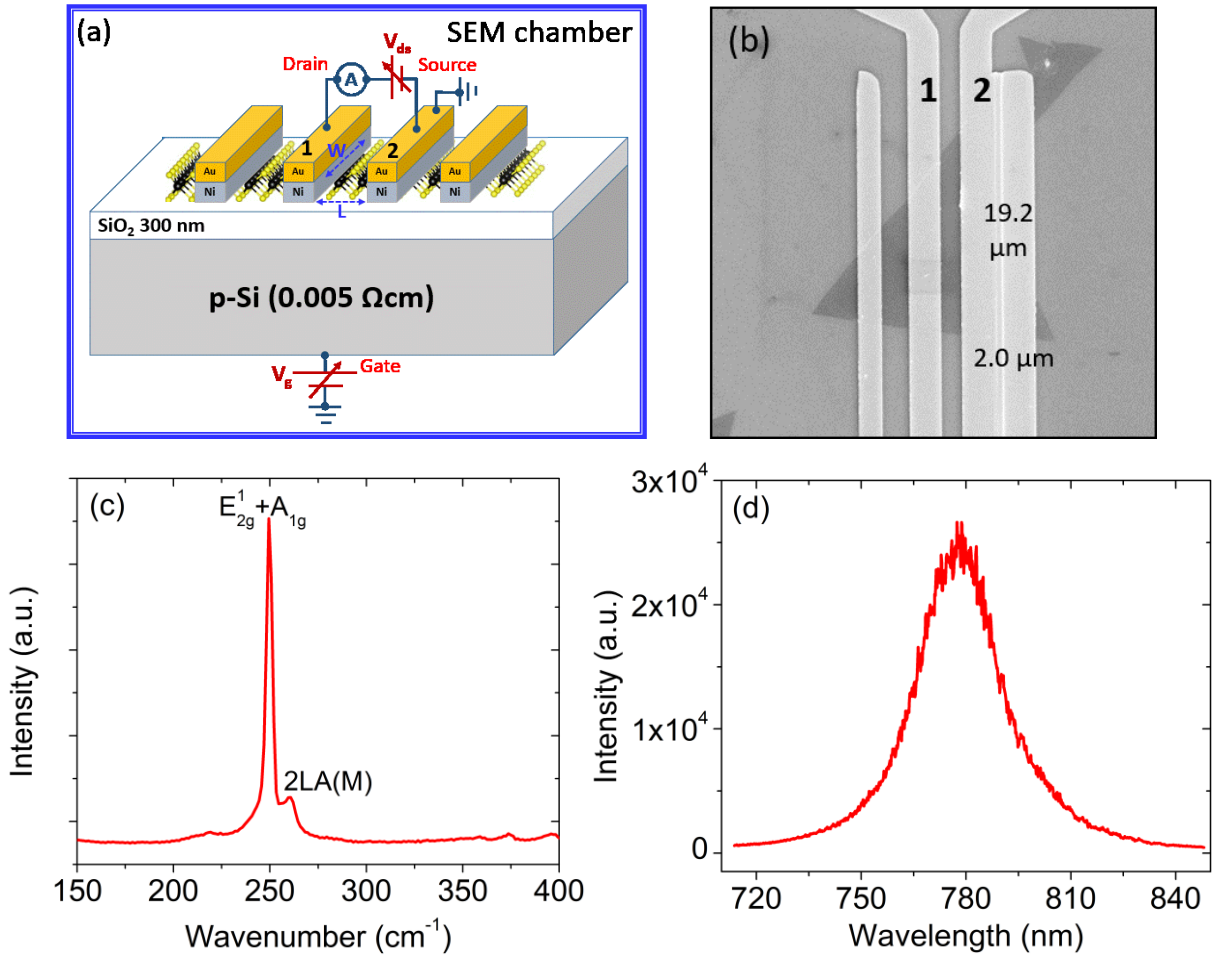

Figure 1 - (a) Layout of a back-gate transistor fabricated with WSe$_2$ on a SiO$_2$/Si substrate. (b) SEM top view of a CVD-grown WSe$_2$ contacted with Ni/Au parallel leads. The device between leads 1 and 2 is used for the electrical measurements. (c) Raman spectrum of the flake under excitation wavelength of 532 nm showing $E_{2g}^1 + A_{1g}$ peak and 2LA(M) peak at frequencies of $\sim 250\ cm^{-1}$ and $\sim 260\ cm^{-1}$, respectively. (d) Photoluminescence spectrum of the flake with maximum at $\sim 778$ nm and FWHM $\sim 23\ cm^{-1}$. Both Raman and PL spectra indicate that the flake is a monolayer.

Fig. 1(a) shows the layout of a typical device, where the Si substrate, functioning as the back-gate, is connected to a voltage generator and the metal leads, constituting the source and the drain of the transistor, are connected to a source-measurement unit (SMU). Fig. 1(b) shows a scanning electron microscope (SEM) top view of a contacted WSe$_2$ flake. The transistor formed between the leads 1 and 2 corresponds to a WSe$_2$ channel with length $L \approx 2\ \mu m$ and average width $W \approx 19\ \mu m$. The Raman spectrum of the flake with unpolarized incident laser, shown in Fig. 1(c), exhibits two peaks around $\sim 250\ cm^{-1}$ and $\sim 260\ cm^{-1}$. The peak observed at

~250 $cm^{-1}$ stems from a combination of the $E_{2g}^1$ and $A_{1g}$ vibrational modes which correspond to in-plane vibrations of W and Se atoms and out-of-plane vibration of Se atoms, respectively. The peak at ~260 cm$^{-1}$ corresponds to the 2LA(M) phonons, a second order resonant Raman mode due to LA phonons at the M point in the Brillouin zone [48,49]. These peaks, under the excitation wavelength of 532 nm, are typical of a WSe$_2$ mono and bilayers [50–52]. Finally, Fig. 1(d) reports the photoluminescence spectrum of the flake with an intense and narrow peak at ~778 $nm$, corresponding to ~1.59 $eV$ bandgap, a value closer to a monolayer than to a bilayer [50,53].

All electrical measurements were performed with the sample kept inside a Zeiss LEO 1530 SEM chamber in high vacuum (pressure <10$^{-6}$ Torr) and at room temperature. Two W-tips, mounted on two piezoelectric-driven arms installed inside the SEM chamber, and the SEM sample holder were electrically connected to a semiconductor parameter analyser (Keithley 4200-SCS) working as source-measurement units (SMUs) for the three-terminal characterization of the device. The setup allowed voltage and current measurements with resolution of 10 µV and $10^{-13}$ $A$, respectively.

*Results and discussions*

*Transistor characterization* - The output characteristics ($I_{ds}$- $V_{ds}$, drain-to-source current-voltage curves) of the WSe$_2$ FET are shown in Fig. 2(a) for different gate voltages, $V_{gs}$. These curves correspond to a maximum current lower than 30 $\frac{nA}{\mu m}$, a device resistance higher than several tens $M\Omega$, and show asymmetry (rectification) between the positive and the negative $V_{ds}$ sweeps as well as a modulation by the gate voltage. The rectification ratio, defined as the current ratio at $\mp 5\ V$, depends on the gate voltage and is higher at positive $V_{gs}$, as reported in Fig. 2(b). We have extensively discussed a similar behaviour in a previous work on MoS$_2$ FETs demonstrating the relevance of the Schottky barriers (SBs) with slightly different heights, formed by the metal lead on the 2D layered channel [54]. The local variation of the barrier height can be caused by defects, unreacted precursor (visible for instance under lead 1 in Fig. 1(b)) or other process residues which may cause spatial inhomogeneity of the Schottky barrier height.

The gate modulation of the channel current at given drain biases, i.e. the $I_{ds} - V_{gs}$ transfer characteristics, are shown in Fig. 2(c), both on linear and logarithmic scale. The transfer curves reveal high current at positive $V_{gs}$ ("on" state) and current suppression at negative $V_{gs}$ ("off" state), thus indicating an n-type conduction with on/off ratio over $10^3$. WSe$_2$ transistors often show ambipolar conduction, with prevailing p-type behaviour, especially for pristine air-exposed WSe$_2$ contacted by high ($> 5.0\ eV$) work-function metals such as Ni [12,24,55]. N-type transistors are fabricated using suitable, low work-function metal contacts, such as Ti, Ag, In, and Al to favour the alignment of the metal Fermi level to the WSe$_2$ conduction band and achieve small n-type Schottky barrier heights [22,46]. Moreover, on untreated WSe$_2$ devices, the carrier type has been also observed to evolve from p-type to ambipolar, and n-type while increasing the WSe$_2$ channel thickness, due to the change in the bandgap of WSe$_2$ and in the carrier band offsets relative to the metal contacts [56,57]. Oxygen and water, which have low kinetic energy for adsorption, are easily deposited on the air exposed WSe$_2$ surface and cause a p-type doping, which further corroborates p-type behaviour in transistors with high work-function metal contacts [30,58]. The desorption of such adsorbates in high vacuum can transform the $WSe_2$ channel from p to n-type and originate the observed high-resistance Schottky Ni contacts. Furthermore, n-type doping of $WSe_2$ can be favoured by positive charges trapped in the $SiO_2$ gate dielectric or at the $WSe_2/SiO_2$ interface [59,60] or by interstitial/substitutional W defects resulting from the growth in a tungsten rich atmosphere, as the prevailing triangular shapes of the CVD flakes seem to indicate [61].

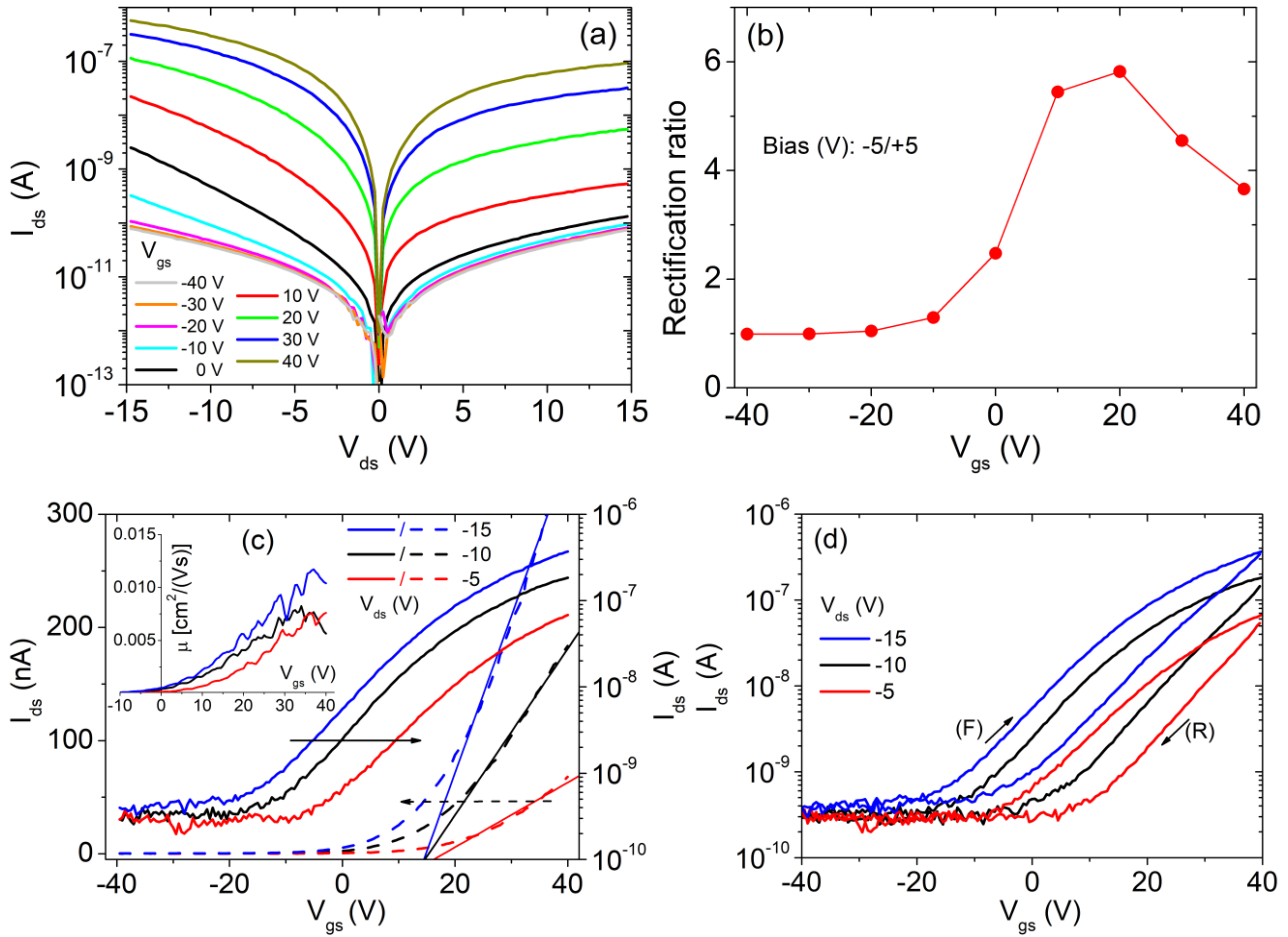

Figure 2 - Electrical characterization of the $WSe_2$ FET. (a) Output characteristics at different gate voltages, showing slight rectifying behaviour. (b) Current rectification ratio at $\mp 5\ V$ ($I_{ds}(-5V)/I_{ds}(+5V)$) as a function of the gate voltage. (c) Transfer characteristics at different drain biases in linear and logarithmic scale; the inset shows the mobility as a function of the gate voltage. (d) Hysteresis in the transfer characteristics during a $V_{gs}$ loop consisting of a reverse and a forward sweep.

The threshold voltage, $V_{th}$, which separates the exponentially increasing off-state current from the on-state current, where the dependence on $V_{gs}$ becomes a linear or power law, can be estimated from the x-axis intersection of the straight lines fitting the transfer curves in the range $V_{gs} > 20\ V$, as shown in Fig. 2(c). Assuming an average $V_{th} \sim 16\ V$ and using the output characteristics, we can evaluate the effective channel mobility [62] as $\mu_{eff} = \frac{L}{W}\frac{1}{C_{ox}}\frac{1}{(V_{gs}-V_{th})}\frac{\partial I_{ds}}{\partial V_{ds}}\Big|_{V_{gs}=cost}$, obtaining a value of $\sim 0.01 \frac{cm^2}{Vs}$ at $V_{gs} = 40\ V$ (here, $L$ and $W$ are the channel length and width, $C_{SiO_2}$ is the capacitance per unit area of the SiO$_2$ gate dielectric: $C_{SiO_2} = \frac{\epsilon_0 \cdot \epsilon_{SiO_2}}{t_{SiO_2}} = 1.15 \cdot 10^{-8} \frac{F}{cm^2}$ with $\epsilon_0$ the vacuum permittivity, $\epsilon_{SiO_2} = 3.9$ and $t_{SiO_2} = 300$ nm the SiO$_2$ thickness). For a more direct comparison to the literature, we can consider the commonly used field-effect channel mobility, $\mu_{FE}$, which uses the transconductance rather than the drain conductance, although its definition is rigorous only for $V_{gs} > V_{th}$ :

$$\mu_{FE} = \frac{L}{W}\frac{1}{C_{ox}}\frac{1}{V_{ds}}\frac{\partial I_{ds}}{\partial V_{gs}}\Big|_{V_{ds}=cost}$$

The inset of Fig. 2(c) shows $\mu_{FE}$ growing linearly with $V_{gs}$ and achieving the value $\mu_{FE} \sim 0.01 \frac{cm^2}{Vs}$ consistent with the estimation of $\mu_{eff}$. The obtained mobility lies on the low side of the range up to 50 $\frac{cm^2}{Vs}$, typically reported for FETs with CVD-grown WSe$_2$ on SiO$_2$ dielectric [6,24,63,64]. The low value of mobility is likely affected by the high contact resistance due to the presence of Schottky barriers; besides, it points toward a defective device with an interfacial trap density, which can be estimated from the subthreshold slope of the transfer curves. Indeed, the sub-threshold swing $SS$, that is the gate voltage change corresponding to one-decade increase of the transistor current, depends on the capacitance per unit area of the trap states, $C_T$, and of the channel depletion layer $C_{DL}$:

$$SS = \frac{dV_{gs}}{d\log I_{ds}} \approx \ln(10)\frac{kT}{q}\left(1 + \frac{C_T + C_{DL}}{C_{SiO_2}}\right)$$

(here, $k$ is the Boltzmann constant, $T$ is the temperature, $q$ is the electron charge). The depletion layer capacitance $C_{DL}$ can be considered null for atomically thin channels, since it can be assumed that the channel is fully depleted below threshold. From the $I_{ds} - V_{gs}$ curves of Fig. 3(c), we obtain a relatively high $SS \sim 15 \frac{V}{decade}$ corresponding to a density of trap states $D_T = \frac{C_T}{q^2} \approx 1.7 \cdot 10^{13}\ cm^{-2}\ eV^{-1}$, a value consistent with similar estimations reported in the literature [65]. The presence of traps, which can be filled or emptied during the forward (F) and reverse (R) gate sweep, manifests as a hysteresis in the transfer characteristics as we have reported and extensively studied for MoS$_2$ back-gated transistors [59]. The width of the hysteresis ($W \sim 11.5$ V at $10^{-8}$ A for the transfer obtained with $V_{ds} = -15\ V$, see Fig. 2(d)) can be used for a rough evaluation of the charge density trapped during a R-F $V_{gs}$ cycle [59]:

$$n_t = W\frac{C_{ox}}{q} = 8.3 \cdot 10^{11} cm^{-2}$$

indicating that less than 5% of the available trap states are filled/emptied during the sweep.

Figs. 2(c) and 2(d) show that threshold voltage, subthreshold swing, mobility and hysteresis width are affected by the applied drain bias. The dependence of the transistor parameters on $V_{ds}$ has been previously studied on similar WSe$_2$ devices and attributed to short channel effects [65]. The $V_{ds}$ dependence, in the high bias regime, is further investigated in Figs. 3(a) and 3(b), which show the transfer curves, the threshold voltage and the off current for increasing drain bias. As pictured in the inset of Fig. 3(b), for a given $V_{gs}$, the increasing absolute value of $V_{ds}$ results in a lowering of the Schottky barrier, $\phi$, and a consequent decrease/rise of the threshold voltage/device current. Such an effect, referred to as drain induced barrier lowering (DIBL), can be quantified as $\left|\frac{\Delta V_{th}}{\Delta V_{ds}}\right| \approx 1.1 \frac{V}{V}$.

Despite the long sweep range, the device does not exhibit a clear transition to p-type conduction (ambipolar behaviour). Here, gate biases higher than 100 V were avoided to prevent device damage.

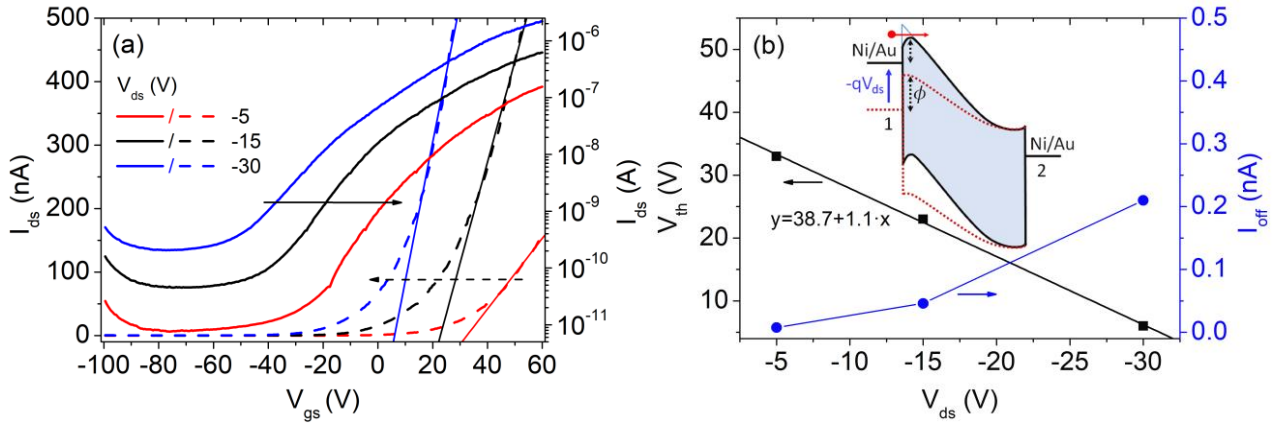

Figure 3 - High-bias electrical characterization of the WSe$_2$ FET. (a) Transfer characteristics at different drain biases in linear and logarithmic scale. (b) Threshold voltage and off current at $V_{gs} = -60\ V$ versus drain bias; the inset shows the drain (1) to source (2) band diagram for increasing drain bias (the brown-dashed curves correspond to lower $V_{ds}$).

*WSe$_2$ vertical field emission transistor* - The n-type gate-tuneable conduction, combined with the favourable geometrical shape and low electron affinity, suggested the investigation of the field emission properties of the WSe$_2$ flakes. The piezoelectric control of the arms inside the SEM chamber, with W-tips movements at a resolution better than $5\ nm$, made it possible to establish the direct W-tip/WSe$_2$ electrical contact. This feature allowed electrical measurements on flakes with useless or missing metal leads. More importantly, we exploited such an opportunity to measure the field emission current from the WSe$_2$ to the W-tip positioned at given distance on the top of the flake. Furthermore, using the SEM sample holder electrically connected to the back-gate of the sample, we were able to observe, for the first time, a gate-modulated FE current from WSe$_2$ monolayer, thus demonstrating the first WSe$_2$ vertical field emission transistor.

The setup used for FE measurements and a SEM image of a measured flake are shown in Figs. 4(a) and 4(b). The flake, which has Raman and PL spectra similar to those reported in Fig. 1, lies partially under a Ni/Au pad, which is used as the source terminal, connected to one of the W-tips. The second W-tip functions as the drain electrode (anode) and can be moved on the flake and accurately positioned at controlled distances $d$ from it. The $I_{ds} - V_{ds}$ curves of Fig. 4(c) are obtained with the drain W-tip in electrical contact with the flake. The similarity with the output characteristics of Fig. 2(a) and the modulation by the gate are evident; moreover, the obvious asymmetry of the contacts enhances the rectifying behaviour and the limited contact area and quality reduces the current. The transfer characteristics of the Fig. 4(d) confirm the n-type behaviour of the $WSe_2$ monolayer and the DIBL effect.

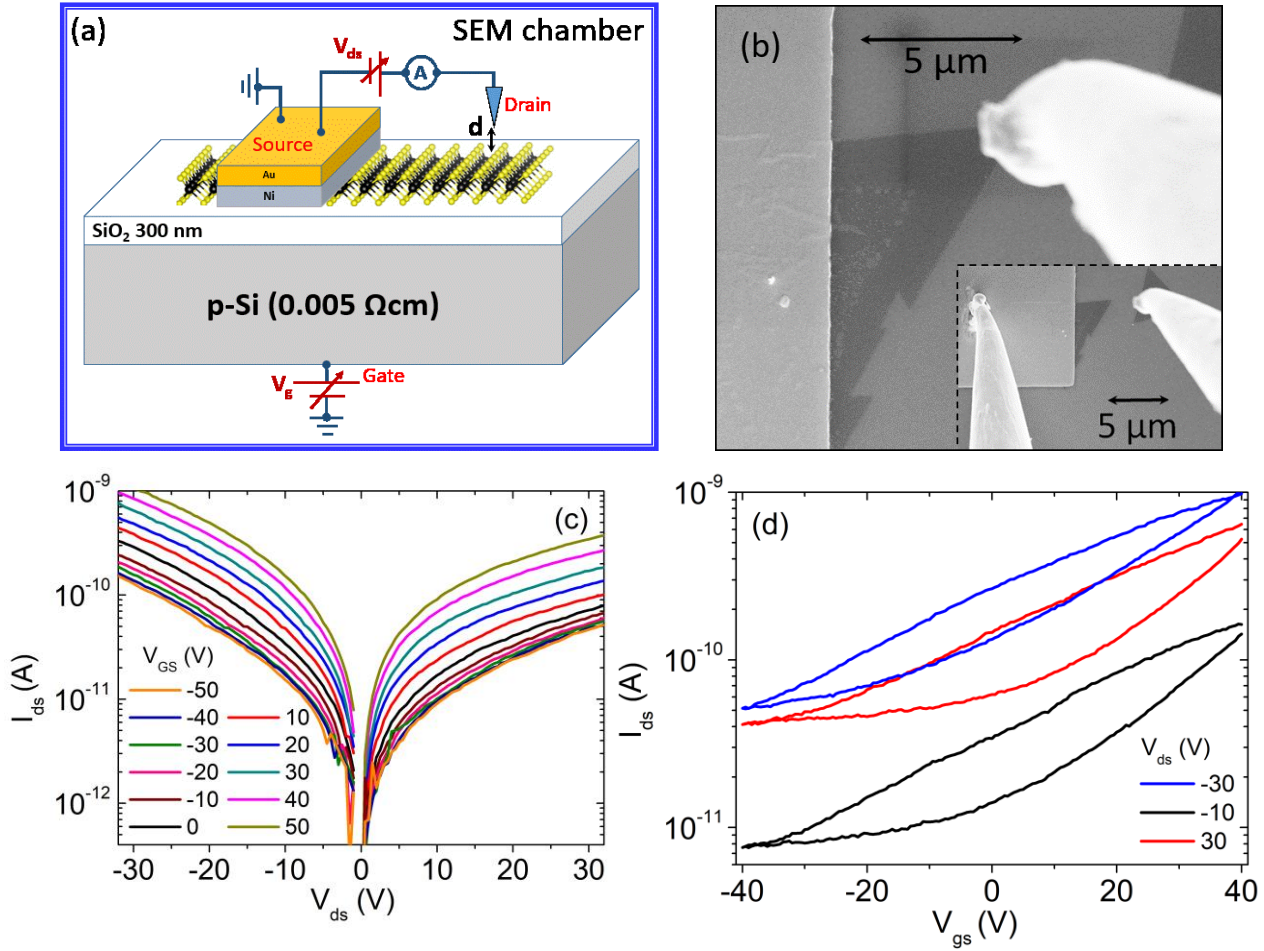

Figure 4 - (a) Layout of a back-gate FE transistor with a $WSe_2$ monolayer channel over a $SiO_2/Si$ substrate. The W-tip labelled as the drain, which collects field emitted electrons and monitors the current, is kept at a distance $d$ from the sample, while the voltage is ramped up. (b) SEM top view of a CVD-grown $WSe_2$ monolayer contacted with a Ni/Au pad (forming the source or cathode) and probed with a W-tip used as the drain (or anode) for the FE current. Output (c) and transfer (d) characteristics of the $WSe_2$ FET with contacts made of a Ni/Au pad and a W-Tip showing a n-type device, strongly affected by drain induced barrier lowering.

The FE current was measured with the drain W-tip at distance $d \sim 400\ nm$ on the top of the flake (Fig. 4(b)). The current was monitored while the voltage was ramped up until 120 V (a constraint imposed to avoid damage of the device or of the measurement setup). The use of a tip-shaped anode is an effective technique to probe the FE of limited areas in alternative to the standard parallel-plate setup that averages the phenomenon over the entire sample [66–70].

Fig. 5(a) shows that, when the W-tip is positioned above the WSe₂ flake and for a drain voltage up to $\sim 60\,V$, the current fluctuates around the noise floor $< 10^{-13} A$, and there is no measurable charge flow. For $V_{ds} > 60\,V$ a steep exponential increase in the current is observed. Although quite noisy, the measured current reasonably follows the behaviour predicted by Fowler-Nordheim (FN) theory of quantum tunnelling through a triangular barrier, as shown in Fig 5(b). According to FN model [71]:

$$I_{ds} = S\, a \frac{E_S^2}{\phi} \cdot exp\left(-b\frac{\phi^{3/2}}{E_S}\right)$$

where S is the emitting surface area, $a = 1.54 \cdot 10^{-6}\, A\,V^{-2} eV$ and $b = 6.83 \cdot 10^7\, V cm^{-1} eV^{-3/2}$ are constants, $E_S\,(Vcm^{-1})$ is the electric field at the emitting surface and $\phi$ is the electron affinity of the emitting material, that is $\sim 3.9\,eV$ for WSe₂ [16,72]. The electric field $E_S = \beta\, V_{ds}/d$, with $\beta$ the so-called field enhancement factor, i.e. the ratio between the electric field at the sample surface and the applied field $V_{ds}/d$. The Fowler-Nordheim model is easily checked though the so-called FN plot of $ln(I_{ds}/V_{ds}^2)$ vs. $1/V_{ds}$, which is expected to exhibit a linear behaviour.

Fig. 5(c) refers to the FE current extracted from a nearby region of the flake with the W-tip at the same distance $d \sim 400\,nm$. Despite the large fluctuations and the dramatic change in the first sweep, it demonstrates a very the important feature, that is the modulation of the FE current by the gate voltage. Indeed, the FE current is enhanced by the application of an increasing positive $V_{gs}$. Fig. 5(d) indicates that the modulation of the FE current by $V_{gs}$ follows an exponential law, with a $SS = \frac{dV_{gs}}{d\log I_{ds}} \sim 33\,\frac{V}{decade}$ comparable to that achieved on the WSe₂ FET discussed before.

We remark that the FE curves typically show large instabilities, with peaks, valleys or permanent changes, due to desorption of adsorbates or atomic-level modification of the emitting surface caused by Joule heating. Hence, some electric conditioning is needed to stabilize the emitting surface before undertaking any systematic FE study. Accordingly, we repeated the FE test after a few trial sweeps and obtained the smoother curves of Fig. 6 (a). Such curves, together with the $I_{ds} - V_{gs}$ transfer characteristic shown in the inset, confirm the exponential voltage-gate modulation of the FE current. Remarkably, the inset of Fig. 6(a) gives a $SS = \frac{dV_{gs}}{d\log I_{ds}} \sim 30\,\frac{V}{decade}$, thus consolidating the previous rough estimation through a direct measurement. We remark that the $SS$ value could be improved by reducing the gate oxide thickness. Assuming a tip diameter of $0.5\,\mu m$, the maximum current of the device attains a value a $\sim 10\,\frac{nA}{\mu m}$, which can be increased by reducing the contact resistance at the source or substantially enhanced by optimizing the field emission extraction process. These observations demonstrate a new device, namely a WSe₂-based vertical field emission transistor.

The FN plot of Fig. 6(b), which proves the FE origin of the observed current, can be used to evaluate the field enhancement factor $\beta = -b\, d\, \phi^{3/2}/m$, where $m$ is the slope of the fitting straight lines. The field enhancement factor results $\sim 55$ and is independent of the gate voltage, as shown in Fig. 6(c). On the contrary, the turn-on field, defined as the field applied to extract a current of 1 pA, shown in the same figure, decreases linearly with $V_{gs}$, with a slope $\frac{dE_{turn-on}}{dV_{gs}} = 0.7\,\frac{V}{\mu m}\frac{1}{V}$ and an average value of $110\,\frac{V}{\mu m}$ over the explored range.

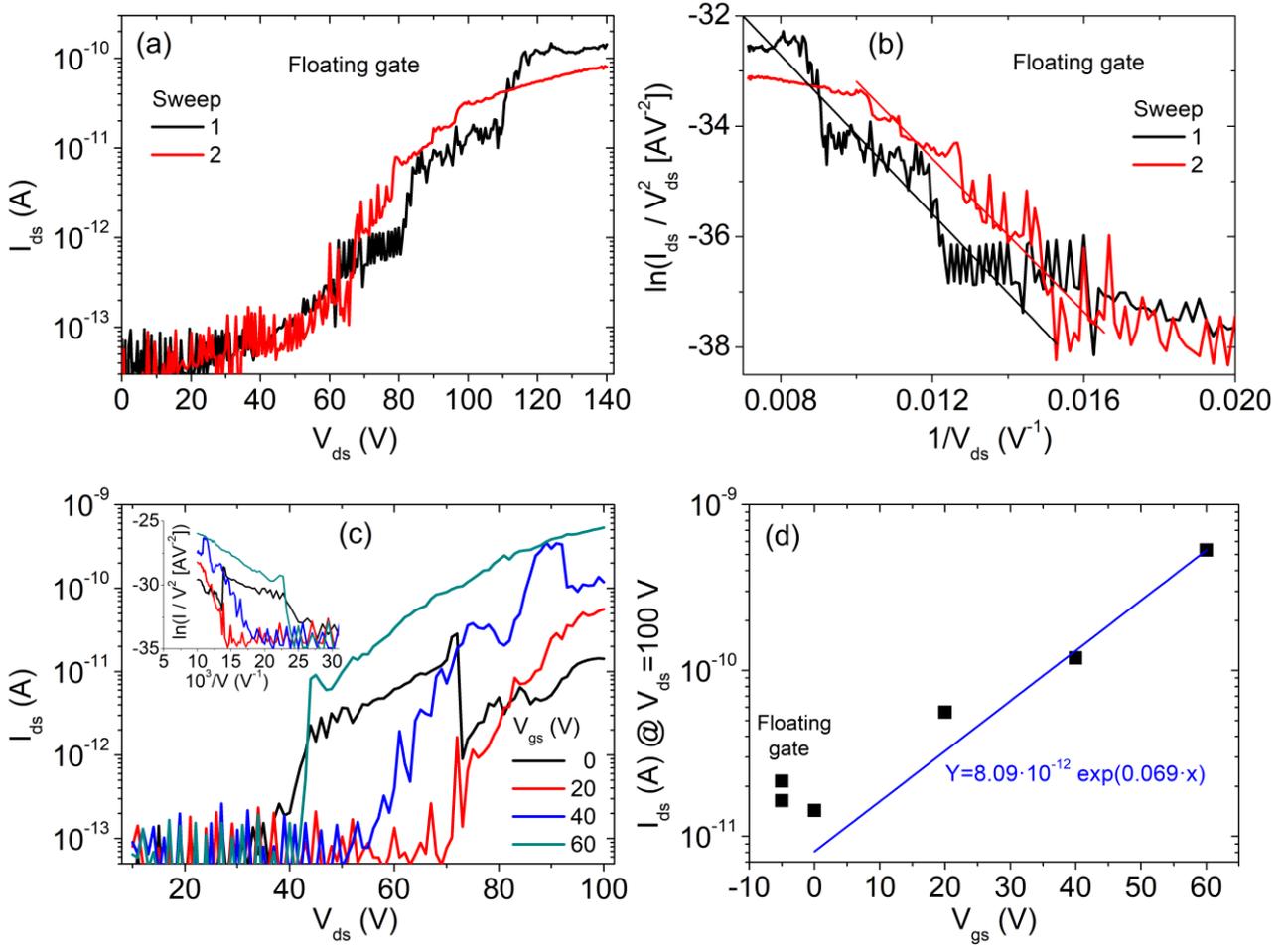

Figure 5 - Field emission measurements with the W-tip at a distance $d \sim 400\ nm$ from the WSe$_2$ flake. (a) FE current measured at floating back-gate and with $V_{ds}$ steps of 0.5 V, and (b) corresponding Fowler-Nordheim plot. (c) FE measurements at given $V_{gs}$ values and $V_{ds}$ steps of 1 V. The corresponding FN plot is shown as inset. (d) Field emission current at $V_{ds} = 100\ V$ showing an exponential dependence on $V_{gs}$ (the values at $V_{gs} = -5\ V$, shown for comparison, refer to floating gate).

We notice that an electric field greater than $10^3\ \frac{V}{\mu m}$ is typically needed to extract electrons from a metal or a semiconductor. The relatively high field or, equivalently, low field enhancement factor (good emitters achieve $\beta \sim 1000$ or more) here obtained are caused by the fact that we have extracted electrons from the flat part of the flake, without taking advantage from the 2D shape of the flake. The geometrical field enhancement would greatly increase the FE current if the extraction was performed from the sharp edge of the flake.

The vertical electron emission from the surface of 2D materials might require a modification of the FN to account for the reduced dimensionality, the energy spectrum, the non-conserving in-plane electron momentum, the finite-temperature and the space-charge-limited effects. A better fit of the experimental data, as shown in the inset of Fig. 6b, can be achieved considering the new high-field regime, $I_{ds} \sim \exp(-b\Phi^{3/2}/E_S)$, of saturated surface field emission proposed for 2D materials.[73]

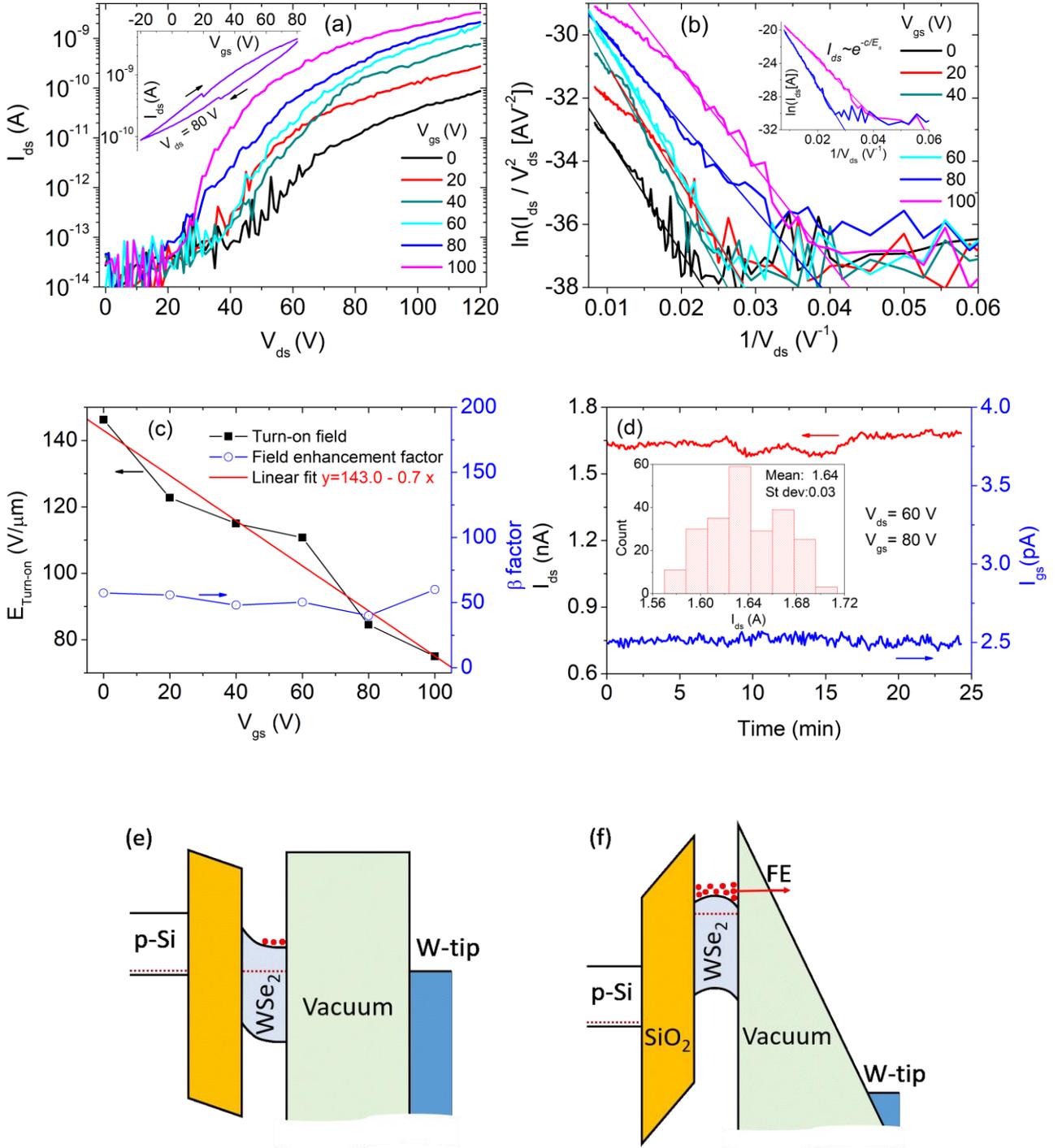

Figure 6 - Field emission measurements after electrical conditioning, with the W-tip at a distance $d \sim 400\ nm$ from the WSe$_2$ monolayer. (a) FE current measured at given back-gate voltages ($V_{ds}$ steps of 1 V), and (b) corresponding Fowler-Nordheim plot (the inset shows the modified F-N plot to consider the effects related to the 2D nature of the emitter). (c) Turn-on field and field enhancement factor versus $V_{gs}$. (d) Field emission

current stability at $V_{ds} = 60\ V$ and $V_{ds} = 80\ V$ (the gate current is monitored as well). (e) Band diagram for (e) the unbiased device and for (f) the device under $V_{ds} > V_{gs} > 0\ V$ bias condition.

Fig. 6 (d) shows that the field emission current has high stability, checked over a time of half an hour, to enable practical applications of the proposed FE transistor.

Finally, Fig. 6 (e) and 6 (f) display the energy band diagram along the vertical direction for the unbiased and biased ($V_{ds} > V_{gs} > 0\ V$) devices and clarify the proposed physical mechanism underlying the operation of the device. The application of a positive gate voltage induces n-doping in the $WSe_2$ channel. The availability of electrons impinging on the field narrowed vacuum barrier, resulting from the $V_{ds}$ voltage, favors the FE current, which is thereby gate-controlled.

In a further development, electrons could be emitted from the edge of the flake which would substantially enhance the drive current capability of the device and lower the operational voltages. Also, a reduced source contact resistance and a thinner gate dielectric would enhance the channel conductance and the transconductance of the device. Obviously, this requires a more complex layout and is the matter for further dedicated work.

*Conclusions*

We have synthesized and characterized $WSe_2$ monolayers on a $SiO_2$/Si substrate, which we have contacted by Ni to form back-gated field effect transistors. We have taken advantage of the gate-controlled n-type conduction of the $WSe_2$ flakes to report, for the first time, their field emission properties. We have shown that electrons can be efficiently extracted from the flat part of a $WSe_2$ monolayer upon the application of an electric field $\sim 100\ \frac{V}{\mu m}$. We have proven that the field emission current can be modulated by the back gate and is stable. We have thereby demonstrated the first vertical field emission transistor suggesting a model for the underlying physics mechanisms. The device can be optimized by reducing the thickness of the gate dielectric to improve its transconductance and by reducing the source contact resistance to increase its current driving capability. The operational bias can be further lowered, and the drive current can be dramatically enhanced by fully exploiting the 2D nature of $WSe_2$ flake, if the extraction of electrons is performed from the edge instead of the flat part in an optimized horizontal device layout.

*Acknowledgments*

We acknowledge the economic support by POR Campania FSE 2014–2020, Asse III Ob. specifico l4, D.D. n. 80 del 31/05/2016 and LR num. 5/2002 Progetti 2008, Prot. 2014, 0293185, 29/04/2014, and CNR-SPIN SEED Project 2017.

*References*

1  D. Jariwala, V. K. Sangwan, L. J. Lauhon, T. J. Marks and M. C. Hersam, *ACS Nano*, 2014, **8**, 1102–1120.
2  A. Gupta, T. Sakthivel and S. Seal, *Progress in Materials Science*, 2015, **73**, 44–126.

3   W. Choi, N. Choudhary, G. H. Han, J. Park, D. Akinwande and Y. H. Lee, *Materials Today*, 2017, **20**, 116–130.
4   A. Jawaid, D. Nepal, K. Park, M. Jespersen, A. Qualley, P. Mirau, L. F. Drummy and R. A. Vaia, *Chem. Mater.*, 2016, **28**, 337–348.
5   S. L. Wong, H. Liu and D. Z. Chi, *Recent progress in chemical vapor deposition growth of two-dimensional transition metal dichalcogenides:*, Progress in Crystal Growth and Characterization of Materials, 2016, vol. 62.
6   J. Huang, L. Yang, D. Liu, J. Chen, Q. Fu, Y. Xiong, F. Lin and B. Xiang, *Nanoscale*, 2015, **7**, 4193–4198.
7   C. R. Serrao, A. M. Diamond, S.-L. Hsu, L. You, S. Gadgil, J. Clarkson, C. Carraro, R. Maboudian, C. Hu and S. Salahuddin, *Appl. Phys. Lett.*, 2015, **106**, 052101.
8   R. Yue, A. T. Barton, H. Zhu, A. Azcatl, L. F. Pena, J. Wang, X. Peng, N. Lu, L. Cheng, R. Addou, S. McDonnell, L. Colombo, J. W. P. Hsu, J. Kim, M. J. Kim, R. M. Wallace and C. L. Hinkle, *ACS Nano*, 2015, **9**, 474–480.
9   L. A. Walsh, R. Addou, R. M. Wallace and C. L. Hinkle, in *Molecular Beam Epitaxy (Second Edition)*, ed. M. Henini, Elsevier, 2018, pp. 515–531.
10  N. Onofrio, D. Guzman and A. Strachan, *Journal of Applied Physics*, 2017, **122**, 185102.
11  F. Giannazzo, G. Fisichella, G. Greco, S. Di Franco, I. Deretzis, A. La Magna, C. Bongiorno, G. Nicotra, C. Spinella, M. Scopelliti, B. Pignataro, S. Agnello and F. Roccaforte, *ACS Appl. Mater. Interfaces*, 2017, **9**, 23164–23174.
12  S. Das and J. Appenzeller, *Appl. Phys. Lett.*, 2013, **103**, 103501.
13  H. Zhou, C. Wang, J. C. Shaw, R. Cheng, Y. Chen, X. Huang, Y. Liu, N. O. Weiss, Z. Lin, Y. Huang and X. Duan, *Nano Lett.*, 2015, **15**, 709–713.
14  S. T. Lee, I. T. Cho, W. M. Kang, B. G. Park and J.-H. Lee, *Nano Convergence*, 2016, **3**, 31.
15  D. Li, X. Wang, Y. Chen, S. Zhu, F. Gong, G. Wu, C. Meng, L. Liu, L. Wang, Tie Lin, S. Sun, H. Shen, X. Wang, W. Hu, J. Wang, J. Sun, X. Meng and J. Chu, *Nanotechnology*, 2018, **29**, 105202.
16  F. A. Rasmussen and K. S. Thygesen, *J. Phys. Chem. C*, 2015, **119**, 13169–13183.
17  A. Prakash and J. Appenzeller, *ACS Nano*, 2017, **11**, 1626–1632.
18  W.-T. Hsu, L.-S. Lu, D. Wang, J.-K. Huang, M.-Y. Li, T.-R. Chang, Y.-C. Chou, Z.-Y. Juang, H.-T. Jeng, L.-J. Li and W.-H. Chang, *Nature Communications*, 2017, **8**, 929.
19  X. Dai, W. Li, T. Wang, X. Wang and C. Zhai, *Journal of Applied Physics*, 2015, **117**, 084310.
20  V. Podzorov, M. E. Gershenson, C. Kloc, R. Zeis and E. Bucher, *Appl. Phys. Lett.*, 2004, **84**, 3301–3303.
21  H.-J. Chuang, B. Chamlagain, M. Koehler, M. M. Perera, J. Yan, D. Mandrus, D. Tománek and Z. Zhou, *Nano Lett.*, 2016, **16**, 1896–1902.
22  W. Liu, J. Kang, D. Sarkar, Y. Khatami, D. Jena and K. Banerjee, *Nano Lett.*, 2013, **13**, 1983–1990.
23  H. C. P. Movva, A. Rai, S. Kang, K. Kim, B. Fallahazad, T. Taniguchi, K. Watanabe, E. Tutuc and S. K. Banerjee, *ACS Nano*, 2015, **9**, 10402–10410.
24  B. Liu, Y. Ma, A. Zhang, L. Chen, A. N. Abbas, Y. Liu, C. Shen, H. Wan and C. Zhou, *ACS Nano*, 2016, **10**, 5153–5160.
25  M. Tosun, S. Chuang, H. Fang, A. B. Sachid, M. Hettick, Y. Lin, Y. Zeng and A. Javey, *ACS Nano*, 2014, **8**, 4948–4953.
26  H. Fang, S. Chuang, T. C. Chang, K. Takei, T. Takahashi and A. Javey, *Nano Lett.*, 2012, **12**, 3788–3792.
27  M. Massicotte, P. Schmidt, F. Vialla, K. G. Schädler, A. Reserbat-Plantey, K. Watanabe, T. Taniguchi, K. J. Tielrooij and F. H. L. Koppens, *Nature Nanotechnology*, 2015, **11**, 42.
28  Z. Wang, Z. Dong, Y. Gu, Y.-H. Chang, L. Zhang, L.-J. Li, W. Zhao, G. Eda, W. Zhang, G. Grinblat, S. A. Maier, J. K. W. Yang, C.-W. Qiu and A. T. S. Wee, *Nature Communications*, 2016, **7**, 11283.
29  D. A. Nguyen, H. M. Oh, N. T. Duong, S. Bang, S. J. Yoon and M. S. Jeong, *ACS Appl. Mater. Interfaces*, 2018, **10**, 10322–10329.
30  Z. Zheng, T. Zhang, J. Yao, Y. Zhang, J. Xu and G. Yang, *Nanotechnology*, 2016, **27**, 225501.
31  J. S. Ross, P. Klement, A. M. Jones, N. J. Ghimire, J. Yan, D. G. Mandrus, T. Taniguchi, K. Watanabe, K. Kitamura, W. Yao, D. H. Cobden and X. Xu, *Nature Nanotechnology*, 2014, **9**, 268.
32  D.-H. Kang, J. Shim, S. K. Jang, J. Jeon, M. Hwan Jeon, G. Young Yeom, W.-S. Jung, Y. Hee Jang, S. Lee and J.-H. Park, *Controllable Nondegenerate p-Type Doping of Tungsten Diselenide by Octadecyltrichlorosilane*, 2015, vol. 9.

33 J. Suh, T.-E. Park, L. Der Yuh, D. Fu, J. Park, H. J. Jung, Y. Chen, C. Ko, C. Jang, Y. Sun, R. Sinclair, J. Chang, S. Tongay and J. Wu, *Doping against the Native Propensity of MoS2: Degenerate Hole Doping by Cation Substitution*, 2014, vol. 14.
34 L. Yang, K. Majumdar, H. Liu, Y. Du, H. Wu, M. Hatzistergos, P. Y. Hung, R. Tieckelmann, W. Tsai, C. Hobbs and P. D. Ye, *Chloride Molecular Doping Technique on 2D Materials: WS2 and MoS2*, 2014, vol. 14.
35 M. Yun, A. Turner, R. J. Roedel and M. N. Kozicki, *Journal of Vacuum Science & Technology B: Microelectronics and Nanometer Structures Processing, Measurement, and Phenomena*, 1999, **17**, 1561–1566.
36 W. P. Kang, J. L. Davidson, A. Wisitsora-at, Y. M. Wong, R. Takalkar, K. Holmes and D. V. Kerns, *Diamond and Related Materials*, 2004, **13**, 1944–1948.
37 J. F. Palma and S. Mil'shtein, in *International Vacuum Nanoelectronics Conference*, 2010, pp. 36–37.
38 G. Wu, X. Wei, S. Gao, Q. Chen and L. Peng, *Nature Communications*, 2016, **7**, 11513.
39 C. Li, X. Yang, S. Ding, Q. Dai, X. Qiu, W. Lei, X. Zhang and B. Wang, *IEEE Electron Device Letters*, 2014, **35**, 786–788.
40 J. L. Shaw, J. B. Boos, B. Kong and J. T. Robinson, in *2018 IEEE International Vacuum Electronics Conference (IVEC)*, 2018, pp. 191–192.
41 A. Di Bartolomeo, F. Giubileo, L. Iemmo, F. Romeo, S. Russo, S. Unal, M. Passacantando, V. Grossi and A. M. Cucolo, *Applied Physics Letters*, 2016, **109**, 023510.
42 R. V. Kashid, D. J. Late, S. S. Chou, Y.-K. Huang, M. De, D. S. Joag, M. A. More and V. P. Dravid, *Small*, 2013, **9**, 2730–2734.
43 F. Giubileo, L. Iemmo, M. Passacantando, F. Urban, G. Luongo, L. Sun, G. Amato, E. Enrico and A. Di Bartolomeo, *arXiv:1808.01185*.
44 F. Urban, M. Passacantando, F. Giubileo, L. Iemmo and A. Di Bartolomeo, *Nanomaterials*, 2018, **8**, 151.
45 R. Addou, C. M. Smyth, J.-Y. Noh, Y.-C. Lin, Y. Pan, S. M. Eichfeld, Stefan Fölsch, J. A. Robinson, K. Cho, R. M. Feenstra and R. M. Wallace, *2D Materials*, 2018, **5**, 025017.
46 W. Liu, W. Cao, J. Kang and K. Banerjee, *(Invited) High-Performance Field-Effect-Transistors on Monolayer-WSe2*, 2013, vol. 58.
47 M. O'Brien, N. McEvoy, T. Hallam, H.-Y. Kim, N. C. Berner, D. Hanlon, K. Lee, J. N. Coleman and G. S. Duesberg, *Scientific Reports*, 2014, **4**, 7374.
48 W. Zhao, Z. Ghorannevis, K. K. Amara, J. R. Pang, M. Toh, X. Zhang, C. Kloc, P. H. Tan and G. Eda, *Nanoscale*, 2013, **5**, 9677–9683.
49 M. O'Brien, N. McEvoy, D. Hanlon, T. Hallam, J. N. Coleman and G. S. Duesberg, *Scientific Reports*, 2016, **6**, 19476.
50 H. Sahin, S. Tongay, S. Horzum, W. Fan, J. Zhou, J. Li, J. Wu and F. M. Peeters, *Phys. Rev. B*, 2013, **87**, 165409.
51 H. Terrones, E. D. Corro, S. Feng, J. M. Poumirol, D. Rhodes, D. Smirnov, N. R. Pradhan, Z. Lin, M. A. T. Nguyen, A. L. Elías, T. E. Mallouk, L. Balicas, M. A. Pimenta and M. Terrones, *Scientific Reports*, 2014, **4**, 4215.
52 H. Zeng, G.-B. Liu, J. Dai, Y. Yan, B. Zhu, R. He, L. Xie, S. Xu, X. Chen, W. Yao and X. Cui, *Scientific Reports*, 2013, **3**, 1608.
53 C. Feng, J. Xiang, P. Liu and B. Xiang, *Materials Research Express*, 2017, **4**, 095703.
54 A. Di Bartolomeo, A. Grillo, F. Urban, L. Iemmo, F. Giubileo, G. Luongo, G. Amato, L. Croin, L. Sun, S. J. Liang and L. K. Ang, *Advanced Functional Materials*, 2018, **28**, 1800657.
55 W.-M. Kang, S. Lee, I.-T. Cho, T. H. Park, H. Shin, C. S. Hwang, C. Lee, B.-G. Park and J.-H. Lee, *Solid-State Electronics*, 2018, **140**, 2–7.
56 C. Zhou, Y. Zhao, S. Raju, Y. Wang, Z. Lin, M. Chan and Y. Chai, *Advanced Functional Materials*, 2016, **26**, 4223–4230.
57 P. R. Pudasaini, A. Oyedele, C. Zhang, M. G. Stanford, N. Cross, A. T. Wong, A. N. Hoffman, K. Xiao, G. Duscher, D. G. Mandrus, T. Z. Ward and P. D. Rack, *Nano Research*, 2018, **11**, 722–730.
58 R. C. Longo, R. Addou, S. KC, J.-Y. Noh, C. M. Smyth, D. Barrera, Chenxi Zhang, J. W. P. Hsu, R. M. Wallace and K. Cho, *2D Materials*, 2017, **4**, 025050.

59 A. Di Bartolomeo, L. Genovese, F. Giubileo, L. Iemmo, G. Luongo, Tobias Foller and M. Schleberger, *2D Materials*, 2018, **5**, 015014.
60 A. Di Bartolomeo, L. Genovese, T. Foller, F. Giubileo, G. Luongo, Luca Croin, S.-J. Liang, L. K. Ang and M. Schleberger, *Nanotechnology*, 2017, **28**, 214002.
61 S. Wang, Y. Rong, Y. Fan, M. Pacios, H. Bhaskaran, K. He and J. Warner, *Shape Evolution of Monolayer MoS 2 Crystals Grown by Chemical Vapor Deposition*, 2014, vol. 26.
62 D. K. Schroder, *Semiconductor Material and Device Characterization*, Wiley, 2006.
63 B. Liu, M. Fathi, L. Chen, A. Abbas, Y. Ma and C. Zhou, *ACS Nano*, 2015, **9**, 6119–6127.
64 S. Li, S. Wang, D.-M. Tang, W. Zhao, H. Xu, L. Chu, Y. Bando, D. Golberg and G. Eda, *Applied Materials Today*, 2015, **1**, 60–66.
65 Z. Yao, J. Liu, K. Xu, E. K. C. Chow and W. Zhu, *Scientific Reports*, 2018, **8**, 5221.
66 A. Di Bartolomeo, A. Scarfato, F. Giubileo, F. Bobba, M. Biasiucci, A. M. Cucolo, S. Santucci and M. Passacantando, *Carbon*, 2007, **45**, 2957–2971.
67 F. Giubileo, A. Di Bartolomeo, A. Scarfato, L. Iemmo, F. Bobba, M. Passacantando, S. Santucci and A. M. Cucolo, *Carbon*, 2009, **47**, 1074–1080.
68 L. Iemmo, A. Di Bartolomeo, F. Giubileo, G. Luongo, M. Passacantando, G. Niu, F. Hatami, O. Skibitzki and T. Schroeder, *Nanotechnology*, 2017, **28**, 495705.
69 F. Giubileo, A. Di Bartolomeo, L. Iemmo, G. Luongo and F. Urban, *Applied Sciences*, 2018, **8**, 526.
70 A. Di Bartolomeo, M. Passacantando, G. Niu, V. Schlykow, G. Lupina, F. Giubileo and T. Schroeder, *Nanotechnology*, 2016, **27**, 485707.
71 Fowler and Nordheim, *Proceedings of the Royal Society of London A: Mathematical, Physical and Engineering Sciences*, 1928, **119**, 173–181.
72 C. M. Smyth, R. Addou, S. McDonnell, C. L. Hinkle and R. M. Wallace, *2D Materials*, 2017, **4**, 025084.
73 Y.S. Ang, M. Zubair, K. J. A. Ooi and L. K. Ang, *arXiv*:1711.05898.